\documentclass[12pt]{iopart}

\usepackage[mode=buildnew]{standalone}
\usepackage{graphicx}
\usepackage{dcolumn}
\usepackage{bm}
\usepackage{braket}
\usepackage{xcolor}
\usepackage[utf8]{inputenc}
\usepackage[T1]{fontenc}
\usepackage[english]{babel}
\usepackage{letltxmacro}
\usepackage{cite}
\usepackage{tikz}
\usetikzlibrary{decorations}
\usetikzlibrary{decorations.pathreplacing} 

\makeatletter
\DeclareRobustCommand{\text}{%
  \ifmmode\expandafter\text@\else\expandafter\mbox\fi}
\let\nfss@text\text
\def\text@#1{{\mathchoice
  {\textdef@\displaystyle\f@size{#1}}%
  {\textdef@\textstyle\f@size{#1}}%
  {\textdef@\textstyle\sf@size{#1}}%
  {\textdef@\textstyle \ssf@size{#1}}%
  \check@mathfonts
  }%
}
\def\textdef@#1#2#3{\hbox{{%
                    \everymath{#1}%
                    \let\f@size#2\selectfont
                    #3}}}
\makeatother

\begin{document}

\title[Entropy production from waiting-time distributions for overdamped dynamics]{Entropy production from waiting-time distributions for overdamped Langevin dynamics}

\author{Ellen Meyberg, Julius Degünther, and Udo Seifert}

\address{II. Institut für Theoretische Physik, Universität Stuttgart, 70550 Stuttgart, Germany}
\ead{meyberg@theo2.physik.uni-stuttgart.de}
\vspace{10pt}
\begin{indented}
\item[]28 February 2024
\end{indented}

%\title{Entropy production from waiting-time distributions \\ for overdamped Langevin dynamics}% Force line breaks with \\

%\author{Ellen Meyberg}
%\author{Julius Degünther}
%\author{Udo Seifert}
%\affiliation{
% II. Institut für Theoretische Physik, Universität Stuttgart, 70550 Stuttgart, Germany
%}

%\date{\today}

\begin{abstract}
For a Markovian dynamics on discrete states, the logarithmic ratio of waiting-time distributions between two successive, instantaneous transitions in forward and backward direction is a measure of time-irreversibility. It thus serves as an entropy estimator, which is exact in the case of a uni-cyclic network. We adopt this framework to overdamped Langevin dynamics, where such transitions have finite duration. By introducing milestones based on the observation of a particle at at least three points, we identify an entropy estimator that becomes exact for driven motion along a one-dimensional potential.
\end{abstract}


\submitto{\jpa}

\maketitle

\textit{Introduction and Motivation.--}
Estimating entropy production from coarse-grained data is a challenging problem in thermodynamic inference \cite{seifert_stochastic_2019}. Recently, waiting-time distributions have been identified as an efficient tool for models with an underlying
Markovian dynamics on a network of discrete states \cite{martinez2019inferring, skinner_estimating_2021, ehrich_tightest_2021, nitzan_universal_2022}. These estimators complement other inequalities, such as  \cite{ barato_thermodynamic_2015, TUR_proof, polettini_effective_2017, stopping_times, speed_limits, pietzonka_thermodynamic_2023, liang_thermodynamic_2023, ohga2023thermodynamic, dechant2023thermodynamic}, that also provide lower bounds on quantifying irreversibility.
%of which the thermodynamic uncertainty relation \cite{barato_thermodynamic_2015, TUR_proof} provides the most prominent example.
Entropy estimators based on waiting-time statistics between few observable transitions in a Markov network have gained particular interest \cite{van_der_meer_thermodynamic_2022, harunari_what_2022}.
%Applying the waiting-time estimator does not require full access of the system but only the measurement of times between successive transitions and the distinction between forward and backward ones. 
There, the time-reversal asymmetry is encoded in the difference between the waiting-time distributions of two successive forward and backward transitions. Surprisingly, these estimators recover the full entropy production rate in a uni-cyclic system with a single observed transition.
%This leads to an estimator that fully quantifies the entropy production in a uni-cyclic system and thus beats other relations like the thermodynamic uncertainty relation \cite{barato_thermodynamic_2015}. 
More generally, the waiting-time approach can be applied to any observations of Markovian events \cite{van_der_meer_time-resolved_2023}, not only to transitions. 
%Another recently examined method to infer time-irreversibility based on the asymmetry of propagators \cite{liang_thermodynamic_2023} is strongly connected to this waiting-time formalism.

Whether and how these concepts can be applied to systems with an underlying continuous dynamics, like an overdamped Langevin dynamics, seems to have been explored less extensively yet. One route is to lump the continuous states into discrete ones \cite{gernert_waiting_2014, esposito, ghosal_inferring_2022} using for example Kramer's rates \cite{Review_kramer}. However, except for limiting cases, state-lumping yields an effective non-Markovian dynamics, which typically prevents recovering the full entropy production \cite{godec_challenges_2023}.
Alternatively, state-like events for overdamped Langevin dynamics can be defined through milestones \cite{hartich_emergent_2021, hartich_violation_2023}. The milestoning scheme relies, in the case of one-dimensional motion, on the identification of certain points as states that characterize the state of the system as long as no other such state is visited. It has been shown that this type of coarse-graining preserves the affinity of a cycle \cite{hartich_emergent_2021}. Although that study provides us with a definition of a state-like event that resembles a Markov state in discrete networks, it lacks a definition of a transition between those. 
%that renders the description accessible within experiments. Can we connect the idea of milestoning with the one of waiting-times between transitions?

In this work, we first explain how transitions can be defined for one-dimensional, overdamped Langevin dynamics based on milestoning. In a second step, we prove that the associated affinity and entropy estimators inspired by \cite{van_der_meer_thermodynamic_2022} are exact. We demonstrate that, in order to measure the affinity of a cycle, an experimentalist needs to observe at least three points on it.

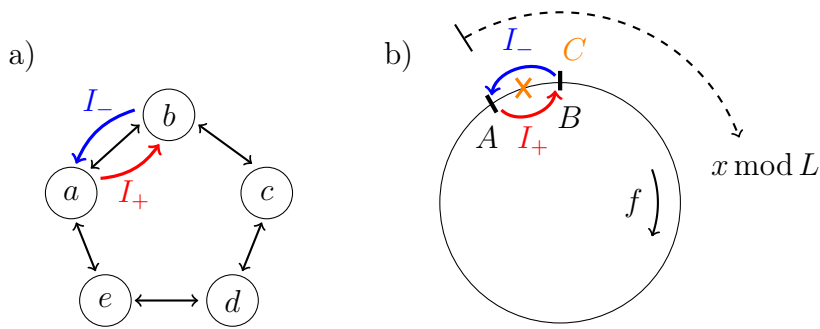
\begin{figure}
\centering
\begin{tikzpicture}[scale=0.65]
\node (A) at (1.0,1.0) {a)};
\node (A) at (8.7,1.0) {b)};
\draw [->, thick,domain=20:-20, samples=100] plot ({2.0*cos(\x)+12}, {2.0*sin(\x)-2});
\node at (13.5,-2.0) {$f$};
\draw[->, dashed, thick,domain=120:20, samples=100] plot ({3.9*cos(\x)+12}, {3.9*sin(\x)-2});
\draw[black, thick] (9.85,1.65) -- (10.2,1.1);
\node (A) at (16.2,-1.2) {$x \, \text{mod} \, L$};
\node (A) at (2,-1.8) {$a$};
\node (B) at (4,-0.2) {$b$};
\node (C) at (6,-1.8) {$c$};
\node (D) at (2.7,-4) {$e$};
\node (E) at (5.3,-4) {$d$};
\draw (2,-1.8) circle (15pt);
\draw (4,-0.2) circle (15pt);
\draw (6,-1.8) circle (15pt);
\draw (2.7,-4) circle (15pt);
\draw (5.3,-4) circle (15pt);
\draw (12,-2) circle (70pt);
\draw[orange, very thick] (11.1,0.15) -- (11.4,0.55);
\draw[orange, very thick] (11.1,0.55) -- (11.4,0.15);
%\filldraw[orange] (11.3,0.30) circle (3pt);
\coordinate[label=above:$A$] (A) at (10.5,-1.1);
\coordinate[label=above:$B$] (B) at (12.2,-0.7);
\coordinate[label=above:$\textcolor{orange}{C}$] (A) at (12.3,0.7);
\draw[black, ultra thick] (10.7,-0.15) -- (10.5,0.2);
\draw[black, ultra thick] (12,0.3) -- (12,0.7);
\draw[->, red, very thick] (2.6,-1.5) to[bend right=25]
node[below] {$I_{+}$} (3.8,-0.8);
\draw[->, blue, very thick] (3.3,-0.1) to[bend right=25]
node[above] {$I_{-}$} (2.1,-1.2);
\draw[->, red, very thick] (10.8,-0.1) to[bend right=65]
node[below] {$I_{+}$} (11.9,0.3);
\draw[->, blue, very thick] (11.9,0.6) to[bend right=65]
node[above] {$I_{-}$} (10.6,0.2);
\draw[<->, black, thick] (2.1,-2.4) -- (2.5,-3.4);
\draw[<->, black, thick] (3.3,-4.0) -- (4.6,-4.0);
\draw[<->, black, thick] (5.5,-3.4) -- (5.9,-2.4);
\draw[<->, black, thick] (5.7,-1.2) -- (4.6,-0.4);
\draw[<->, black, thick] (2.4,-1.3) -- (3.4,-0.4);
\end{tikzpicture}
\caption{Illustration of the analogy between transitions in discrete and continuous one-dimensional stochastic systems. a) Discrete five-state ring with transitions $I_{\pm}$ between states $a$ and $b$, black arrows denote that two states are connected. b) Continuous ring of length $L$ with driving force $f$ and milestones $A$ and $B$, the $(x \, \text{mod} \, L)$-coordinate of state $A$ is arbitrarily set to zero. Observing the additional state $C$ marked with a cross allows to distinguish clockwise and counter clockwise motion between $A$ and $B$.}
\label{fig:ABdiskkont}
\end{figure}

\textit{Model.--}
We consider one-dimensional overdamped Langevin dynamics of a particle on a ring of length $L$ in a periodic, time-independent potential $V(x)$ under the influence of a driving force $f$ described by
\begin{eqnarray}
\dot{x}(t) = \mu F(x) + \xi(t) = \mu[-\partial_x V(x) + f] + \xi(t) \, .
\label{eq:Langevin}
\end{eqnarray}
Here $\mu$ denotes the mobility, the thermal energy $k_{\text{B}}T$ is set to unity and the random force obeys
\begin{eqnarray}
\langle \xi(t) \rangle = 0 \quad \text{and} \quad \langle \xi(t') \xi(t) \rangle = 2 \mu \delta(t-t') \, .
\end{eqnarray}
Equivalently, the dynamics can be described with the Fokker-Planck equation
\begin{eqnarray}
\partial_t p(x,t) = - \partial_x j(x,t) = - \mu \partial_x [F(x) - \partial_x] p(x,t) \, ,
\label{eq:FokkerPlanck}
\end{eqnarray}
which describes the evolution of the probability density $p(x,t)$ with periodic boundary conditions. In the steady state, i.e., for $\partial_t p^{\text{s}}(x) = 0$, a constant stationary current $j^{\text{s}}$ develops.
The driving affinity, i.e., the entropy production per cycle, is given by
\begin{eqnarray}
\mathcal{A} = f L
\label{eq:affinity}
\end{eqnarray}
independently of $V(x)$ leading to the entropy production rate in the stationary state
\begin{eqnarray}
\sigma = j^{\text{s}} \mathcal{A} \, .
\label{eq:EPR}
\end{eqnarray}
For a uni-cyclic, discrete network as shown in figure \ref{fig:ABdiskkont}\,a) the affinity is given by
\begin{eqnarray}
\mathcal{A} = \sum_{(mn)} \ln \left(\frac{k_{mn}}{k_{nm}} \right) \, ,
\end{eqnarray}
where the transition rates $k_{mn}$ and $k_{nm}$ connect two neighbouring states $m$ and $n$. The summation includes all edges $(mn)$ of the cycle. It has been shown in \cite{van_der_meer_thermodynamic_2022} that $\mathcal{A}$ can be inferred from waiting time distributions
\begin{eqnarray}
\psi_{I \rightarrow J}(\tau) \equiv P(J, \tau|I,0) \, ,
\label{eq:WT_CP}
\end{eqnarray}
which is the probability for transition $J$ to happen after a time $\tau$ given the previous transition $I$.  To be specific, we consider the pair of transitions between states $a$ and $b$ called $I_+$ and $I_-$ in figure \ref{fig:ABdiskkont}\,a). The logarithmic ratio of waiting-time distributions
\begin{eqnarray}
\hat{\mathcal{A}}(\tau) \equiv \ln \left(\frac{\psi_{I_{+} \rightarrow I_{+}}(\tau)}{\psi_{I_{-} \rightarrow I_{-}}(\tau)} \right) 
\label{eq:estimator_1}
\end{eqnarray}
between two successive transitions in clockwise, $I_+ \rightarrow I_+$, and anti clockwise direction, $I_- \rightarrow I_-$, serves as time-independent and exact estimator of the affinity $\mathcal{A}$. The proof of this relation crucially depends on the Markovianity of the dynamics \cite{van_der_meer_thermodynamic_2022}. Our goal is to find an equivalent procedure for continuous states as required for the particle on a ring. 

First of all, we need to identify transitions, whose crucial property is to resolve the particle's direction of motion \cite{godec_challenges_2023}. Furthermore, if we can observe only one pair of transitions, two successive ones in the same direction should be equal to a full passage of the cycle.
Since the velocity for overdamped dynamics is not well defined, a definition of a transition based on the directed crossing of a milestone, i.e., a fixed position on the ring, is not possible. 

%Nevertheless, a definition that retains the continuous degree of freedom is required rather than lumping the $x$-coordinate into discrete states. 
%Hence, we need to find an appropriate set of milestones that are equal to discrete points on the $x$-coordinate. The last visited milestone is equal to the state of the system. If there is a second milestone positioned next to a first one in such a way that fast recrossing is prevented, the direction of the transition between the two can be resolved. 
%All considerations made so far lead to the following procedure illustrated in figure \ref{fig:ABdiskkont}b).

We, therefore, have to introduce two milestones $A$ and $B$ through two points with $x$-coordinates
\begin{eqnarray}
A \equiv (x = x_A \; \text{mod} \; L) \quad \text{and} \quad B \equiv (x = x_B \; \text{mod} \; L)
\end{eqnarray}
with periodic boundary conditions as illustrated in figure \ref{fig:ABdiskkont}\,b). Without loss of generality, $x_A$ can be set to zero. If we apply the milestoning scheme as coarse-graining, an effective trajectory thus becomes binary. It is either in state $A$ or $B$ depending on whether the position $x_A$ or $x_B$ was visited latest, respectively. Transitions $I_{\pm}$ are identified as passages in opposite direction between the two milestones,
\begin{eqnarray}
I_+ \equiv A \stackrel{C}{\longrightarrow} B \quad \text{and} \quad I_- \equiv B \stackrel{C}{\longrightarrow} A \, ,
\end{eqnarray}
where $C$ on top of the arrow denotes an auxiliary, third point. The latter ensures that a transition corresponds to the passage along the shorter route between $A$ and $B$, as illustrated in figure \ref{fig:ABdiskkont}\,b). This implies that inference based on waiting times for overdamped Langevin dynamics requires a set of at least three observed points. 
A transition $I_+$ and $I_-$ ends if endpoints $B$ and $A$, respectively, are reached. Crucially, the state of the system is fully characterized each time a transition is finished, because reaching a milestone $A$ or $B$ is a Markovian event \cite{van_der_meer_time-resolved_2023}. In contrast to transitions between discrete states, the continuous ones possess a finite duration, which ends when the milestone is reached. 
%In addition, there exist many different microscopic realizations of a certain transition.

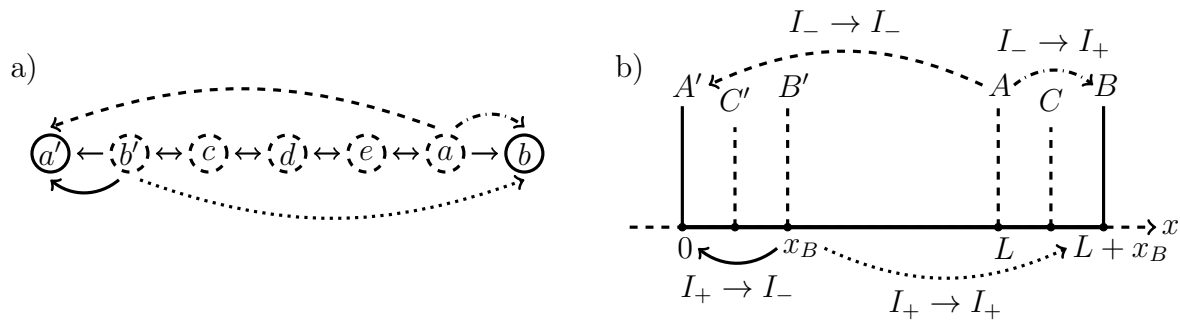
\begin{figure}[h!]
\begin{tikzpicture}[scale=0.7]
\node (A) at (12,5.0) {b)};
\draw[black, ultra thick] (13,2) -- (21,2);
\draw[->, dashed, very thick] (12,2) -- (22,2);
\draw[black, very thick] (13,2) -- (13,4.3);
\draw[black, dashed, very thick] (15,2) -- (15,4.3);
\draw[black, dashed, very thick] (14,2) -- (14,3.9);
\draw[black, dashed, very thick] (20,2) -- (20,3.9);
\draw[black, dashed, very thick] (19,2) -- (19,4.3);
\draw[black, very thick] (21,2) -- (21,4.3);
\filldraw[black] (13,2) circle (2pt);
\filldraw[black] (15,2) circle (2pt);
\filldraw[black] (19,2) circle (2pt);
\filldraw[black] (21,2) circle (2pt);
\filldraw[black] (14,2) circle (2pt);
\filldraw[black] (20,2) circle (2pt);
\coordinate[label=right:$x$] (x) at (21.9,2.0);
\coordinate[label=right:$A'$] (A1) at (12.6,4.7);
\coordinate[label=right:$B'$] (B3) at (14.6,4.7);
\coordinate[label=right:$A$] (A2) at (18.6,4.7);
\coordinate[label=right:$B$] (B4) at (20.6,4.7);
\coordinate[label=right:$0$] (A) at (12.7,1.6);
\coordinate[label=right:$x_B$] (B1) at (14.7,1.6);
\coordinate[label=right:$L$] (A) at (18.7,1.6);
\coordinate[label=right:$L+x_B$] (B2) at (20.2,1.6);
\coordinate[label=above:$C'$] (C) at (14.0,4.0);
\coordinate[label=above:$C$] (C2) at (20.0,4.0);
\draw[->, dotted, very thick] (15.7,1.6) to[bend right=25]
node[below] {$I_{+} \rightarrow I_{+}$} (20.3,1.6);
\draw[->, dashed, very thick] (18.7,4.7) to[bend right=25]
node[above] {$I_- \rightarrow I_-$} (13.5,4.7);
\draw[->, dash dot, very thick] (19.3,4.7) to[bend right=-40]
node[above] {$I_- \rightarrow I_+$} (20.8,4.7);
\draw[->, very thick] (14.8,1.6) to[bend right=-40]
node[below] {$I_+ \rightarrow I_-$} (13.3,1.6);
\node (A) at (0.5,5.0) {a)};
\draw[dashed, very thick] (2.5,3.4) circle (10pt);
\draw[dashed, very thick] (4.0,3.4) circle (10pt);
\draw[dashed, very thick] (5.5,3.4) circle (10pt);
\draw[dashed, very thick] (7.0,3.4) circle (10pt);
\draw[dashed, very thick] (8.5,3.4) circle (10pt);
\draw[very thick] (10.0,3.4) circle (10pt);
\draw[very thick] (1.0,3.4) circle (10pt);
\node (A) at (2.5,3.4) {$b'$};
\node (B) at (4.0,3.4) {$c$};
\node (C) at (5.5,3.4) {$d$};
\node (D) at (7.0,3.4) {$e$};
\node (E) at (8.5,3.4) {$a$};
\node (E) at (10.0,3.4) {$b$};
\node (E) at (1.0,3.4) {$a'$};
\draw[<-, black, thick] (1.5,3.4) -- (2.0,3.4);
\draw[<->, black, thick] (3.0,3.4) -- (3.5,3.4);
\draw[<->, black, thick] (4.5,3.4) -- (5.0,3.4);
\draw[<->, black, thick] (6.0,3.4) -- (6.5,3.4);
\draw[<->, black, thick] (7.5,3.4) -- (8.0,3.4);
\draw[->, black, thick] (9.0,3.4) -- (9.5,3.4);
\draw[->, dotted, very thick] (2.7,2.9) to[bend right=20] (10.0,2.9);
\draw[->, very thick] (2.3,2.9) to[bend right=-40]
(1.0,2.9);
\draw[->, dashed, very thick] (8.3,3.9) to[bend right=20] (1.0,3.9);
\draw[->, dash dot, very thick] (8.7,3.9) to[bend right=-40] (10.0,3.9);
\end{tikzpicture}
\caption{System with absorbing states. a) Discrete network with absorbing states $a'$ and $b$ marked with full circles, otherwise dashed. b) Full vertical lines denote absorbing boundaries placed at milestones $A'$ and $B$. The states $A, A'$, $B, B'$ and $C, C'$ are equivalent points on the ring mod $L$ where we have set $x_A = 0$. The precise location of $C$ between $A$ and $B$ is arbitrary.}
\label{fig:AB-Modell}
\end{figure}

In the discrete case, waiting-time distributions can be determined from solving the absorbing master equation \cite{van_der_meer_thermodynamic_2022, sekimoto2022derivation}, where the observed links lead to states that the particle cannot leave, i.e., to absorbing states, as shown in figure \ref{fig:AB-Modell}\,a). Analogously, in the present continuous case, we have to consider an interval $0 \leq x \leq L + x_B$, as illustrated in figure \ref{fig:AB-Modell}\,b), and to solve the Fokker-Planck equation (\ref{eq:FokkerPlanck}) with absorbing boundary conditions that read \cite{risken_fokker-planck_1996, goel2013stochastic}
\begin{eqnarray}
p(x=0,\tau) = 0 \quad \text{and} \quad p(x = L+x_B, \tau) = 0 \, .
\end{eqnarray}
Waiting-time distributions are thus equal to the fluxes through absorbing boundaries at $A'$ and $B$ given the initial position as fixed by the position after the last transition, here set to $\tau = 0$. This leads to
\numparts
\begin{eqnarray}
\psi_{I_{+} \rightarrow I_{+}}(\tau) = j(L+x_B, \tau|x_B, 0) \, , \\
\psi_{I_{-} \rightarrow I_{-}}(\tau) = -j(0,\tau|L, 0) \, , \\
\psi_{I_{+} \rightarrow I_{-}}(\tau) = -j(0,\tau|x_B, 0) \, , \\
\psi_{I_{-} \rightarrow I_{+}}(\tau) = j(L+x_B,\tau|L, 0)  \, ,
\label{eq:WTD_Fluxes}
\end{eqnarray}
\endnumparts
where the signs are chosen so that the waiting-time distributions are positive. We have conditioned the fluxes on the initial position $x=x_B$ after $I_+$ and $x=L$ after $I_-$, respectively. 
%The expression (\ref{eq:WTD_Fluxes}) can be used to calculate the waiting-time distributions.

\textit{Entropy estimator.--}
After identifying waiting-time distributions for continuous stochastic systems we now want to apply the estimator (\ref{eq:estimator_1}). Without any driving force $f$, i.e. in equilibrium, the cycle is equally likely passed in both directions. However, for non-vanishing $f$ in clockwise-direction, two subsequent $I_+$ transitions get more likely than two $I_-$ ones. Hence, we expect that $\hat{\mathcal{A}}(\tau)$ measures time-irreversibility and is related to the affinity $\mathcal{A}$ of a cycle similar to the discrete case. In fact, we will show that
\begin{eqnarray}
\hat{\mathcal{A}}(\tau) = \mathcal{A} = fL\, .
\label{eq:exact}
\end{eqnarray}
On a theoretical level, waiting-time distributions can also be understood as 
\begin{eqnarray}
\psi_{I \rightarrow J}(\tau) = \sum_{\zeta \in \set{\zeta_{I \rightarrow J}^{\tau}}} \mathcal{P}[\zeta |I] = \sum_{\zeta \in \set{\zeta_{I \rightarrow J}^{\tau}}} \mathcal{P}[\zeta |X_I] \, ,
\label{eq:WTDausPW}
\end{eqnarray}
i.e., as the sum over all path-weights $\mathcal{P}$ of trajectories $\zeta_{I \rightarrow J}^{\tau}$ that start at the endpoint of transition $I$ and that end at the one of $J$ after time $\tau$ without undergoing any other transitions in between. The conditioning on the first transition $I$ in (\ref{eq:WTDausPW}) can be replaced by a conditioning on the milestone $X_I \in \set{A,B}$ at the end of this transition. In the following, we will call microscopic paths connecting two transitions snippets. 

\begin{figure}
\begin{center}
\begin{tikzpicture}[scale=0.6]
\node (A) at (-0.5,9.7) {a)};
\draw[->, dashed, black, thick] (0,9.0) -- (13.0,9.0);
\coordinate[label=right:$t$] (A1) at (13.0,9.0);
\draw[->, black, thick] (0.3,8.1) -- (1.3,8.1);
\draw[->, black, thick] (1.9,8.1) -- (3.4,8.1);
\draw[->, black, thick] (4.0,8.1) -- (7.8,8.1);
\draw[->, black, thick] (8.4,8.1) -- (10.7,8.1);
\draw[->, black, thick] (11.3,8.1) -- (12.8,8.1);
\coordinate[label=above:$B$] (A1) at (-0.1,7.7);
\coordinate[label=above:$A$] (A1) at (1.6,7.7);
\coordinate[label=above:$B$] (A1) at (3.7,7.7);
\coordinate[label=above:$A$] (A1) at (8.1,7.7);
\coordinate[label=above:$B$] (A1) at (11.0,7.7);
\coordinate[label=above:$A$] (A1) at (13.1,7.7);
\node (A) at (0.7,7.4) {$I_-$};
\node (A) at (2.5,7.4) {$I_+$};
\node (A) at (9.5,7.4) {$I_+$};
\node (A) at (12.1,7.4) {$I_-$};
\node (A) at (-0.5,6.6) {b)};
\draw[->, black, very thick] (0,-2) -- (13,-2);
\draw[->, black, very thick] (0,-2) -- (0,5);
\draw[black, dashed, thick] (0,-1) -- (13,-1);
\draw[black, dashed, thick] (0,0) -- (13,0);
\draw[black, dashed, thick] (0,3) -- (13,3);
\draw[black, dashed, thick] (0,4) -- (13,4);
\draw[black, dashed, thick] (12.3,5) -- (12.3,-2);
\draw[black, dashed, thick] (3.7,5) -- (3.7,-2);
\draw[black, dashed, thick] (1.6,5) -- (1.6,-2);
\draw[black, dashed, thick] (2.6,5) -- (2.6,-2);
\draw[black, dashed, thick] (13,5) -- (13,-2);
\draw[black, dashed, thick] (11,5) -- (11,-2);
\draw[black, dashed, thick] (8.1,5) -- (8.1,-2);
\draw[black, very thick] (0,0) -- (0.5,-0.5);
\draw[black, very thick] (0.5,-0.5) -- (0.9,-0.2);
\draw[black, very thick] (0.9,-0.2) -- (1.5,-0.7);
\draw[black, very thick] (1.5,-0.7) -- (1.75,-1.4);
\draw[black, very thick] (1.75,-1.4) -- (2.6,-1);
\draw[black, very thick] (2.6,-1) -- (2.8,-0.6);
%\draw[black, very thick] (2.5,-0.5) -- (2.8,-0.6);
\draw[black, very thick] (2.8,-0.6) -- (3.7,0);
\draw[black, very thick] (3.7,0) -- (4.0,-0.2);
\draw[black, very thick] (4.0,-0.2) -- (5.0,0.5);
\draw[black, very thick] (5.0,0.5) -- (5.5,0.6);
\draw[black, very thick] (5.5,0.6) -- (5.8,1.2);
\draw[black, very thick] (5.8,1.2) -- (6.1,0.9);
\draw[black, very thick] (6.1,0.9) -- (7.2,2.0);
\draw[black, very thick] (7.2,2.0) -- (7.5,2.8);
%\draw[black, very thick] (7.5,2.8) -- (8.1,3.0);
%\draw[black, very thick] (8.1,3.0) -- (8.8,3.7);
%\draw[black, very thick] (8.8,3.7) -- (10,3.4);
%\draw[black, very thick] (10,3.4) -- (11.1,4.9);
\draw[black, very thick] (11.0,4.0) -- (11.5,4.6);
%\draw[black, very thick] (11.1,4.9) -- (11.5,4.3);
\draw[black, very thick] (11.5,4.6) -- (12.3,4);
\draw[black, very thick] (12.3,4) -- (12.5,3.5);
\draw[black, very thick] (12.5,3.5) -- (13,3);
\coordinate[label=right:$A'$] (A1) at (-1.1,-1);
\coordinate[label=right:$B'$] (B1) at (-1.1,0);
\coordinate[label=right:$A$] (A2) at (-1.0,3);
\coordinate[label=right:$B$] (B2) at (-1.0,4);
\coordinate[label=right:$t$] (t) at (13.1,-2);
\coordinate[label=right:$t_{A}$] (tA) at (7.9,-2.4);
\coordinate[label=right:$0$] (t0) at (3.5,-2.4);
\coordinate[label=right:$t_{++}$] (ts) at (10.7,-2.4);
\coordinate[label=left:$x$] (x) at (-0.2,4.8);
%\coordinate[label=right:$-$] (m1) at (0.7,-0.7);
%\coordinate[label=right:$+$] (p1) at (3,-0.5);
%\coordinate[label=right:$+$] (m2) at (9,3.2);
%\coordinate[label=right:$-$] (p2) at (11.6,3.5);
\filldraw[red] (0,0) circle (2pt);
\filldraw[red] (1.6,-1) circle (2pt);
\filldraw[red] (2.6,-1) circle (2pt);
\filldraw[red] (3.7,0) circle (2pt);
\filldraw[red] (8.1,3) circle (2pt);
\filldraw[red] (11.0,4) circle (2pt);
\filldraw[red] (12.3,4) circle (2pt);
\filldraw[red] (13,3) circle (2pt);
%\draw[->, red!50, ultra thick] (3.2,-2.4) to[bend right=25]
%node[below] {$++$} (B2);
%\draw[decorate, yshift=2ex]  (0,0) -- node[above=0.4ex] {$T(-+)$}  (2,0);
%\draw[brace] (0,-2) -- (2.5,-2)
\draw[decorate, decoration={brace,amplitude=6pt}]  (-0.8,5.3) -- node[above=1.2ex] {$\tau_{0}$}  (1.6,5.3);
\filldraw[white] (-0.8,5.3) circle (10pt);
\coordinate[label=right:...] (A2) at (-1.4,5.5);
\draw[decorate, decoration={brace,amplitude=6pt}]  (1.6,5.3) -- node[above=1.2ex] {$\tau_{1}$}  (3.7,5.3);
\draw[decorate, decoration={brace,amplitude=6pt}]  (3.7,5.3) -- node[above=1.2ex] {$\tau_{2}$}  (11.0,5.3);
\draw[decorate, decoration={brace,amplitude=6pt}]  (11.0,5.3) -- node[above=1.2ex] {$\tau_{3}$}  (13,5.3);
\draw[decorate, decoration={brace,amplitude=6pt}]  (8.1,-2.7) -- node[below=1.2ex] {$\tilde{\tau}_{2}$}  (2.6,-2.7);
\draw[decorate, decoration={brace,amplitude=6pt}]  (2.6,-2.7) -- node[below=1.2ex] {$\tilde{\tau}_{3}$}  (0.0,-2.7);
\draw[decorate, decoration={brace,amplitude=6pt}]  (12.3,-2.7) -- node[below=1.2ex] {$\tilde{\tau}_{1}$}  (8.1,-2.7);
\draw[decorate, decoration={brace,amplitude=6pt}]  (14.4,-2.7) -- node[below=1.2ex] {$\tilde{\tau}_{0}$}  (12.3,-2.7);
\filldraw[white] (14.4,-2.7) circle (12pt);
\node (A) at (14.4,-2.85) {...};
\draw[red, dashed, very thick] (3.7,0) -- (4.0,-0.2);
\draw[red, very thick] (4.0,-0.2) -- (5.0,0.5);
\draw[red, very thick] (5.0,0.5) -- (5.5,0.6);
\draw[red, very thick] (5.5,0.6) -- (5.8,1.2);
\draw[red, very thick] (5.8,1.2) -- (6.1,0.9);
\draw[red, very thick] (6.1,0.9) -- (7.2,2.0);
\draw[red, very thick] (7.2,2.0) -- (7.5,2.8);
\draw[red, very thick] (7.5,2.8) -- (8.1,3.0);
\draw[red, dashed, very thick] (8.1,3.0) -- (9.2,3.7);
\draw[red, dashed, very thick] (9.2,3.7) -- (10.3,3.4);
\draw[red, dashed, very thick] (10.3,3.4) -- (11.0,4.0);
\draw[black, very thick] (3.7,0) -- (4.0,-0.2);
\draw[black, very thick] (4.0,-0.2) -- (5.0,0.5);
\draw[black, very thick] (5.0,0.5) -- (5.5,0.6);
\draw[black, very thick] (5.5,0.6) -- (5.8,1.2);
\draw[black, very thick] (5.8,1.2) -- (6.1,0.9);
\draw[black, very thick] (6.1,0.9) -- (7.2,2.0);
\draw[black, very thick] (7.2,2.0) -- (7.5,2.8);
\draw[black, very thick] (7.5,2.8) -- (8.1,3.0);
\draw[red, dashed, very thick] (8.1,-1.0) -- (9.2,-0.3);
\draw[red, dashed, very thick] (9.2,-0.3) -- (10.3,-0.6);
\draw[red, dashed, very thick] (10.3,-0.6) -- (11.0,0.0);
\filldraw[red] (8.1,-1) circle (2pt);
\filldraw[red] (11.0,0) circle (2pt);
\draw[->, red, very thick] (9.3,2) -- (9.3,1);
\draw[->, red, very thick] (6.5,-0.5) -- (5.5,-0.5);
\draw[red, dashed, very thick] (0.7,-1.0) -- (1.8,-0.3);
\draw[red, dashed, very thick] (1.8,-0.3) -- (2.9,-0.6);
\draw[red, dashed, very thick] (2.9,-0.6) -- (3.6,0.0);
\filldraw[red] (0.7,-1.0) circle (2pt);
\node (A) at (-0.5,-4.0) {c)};
\draw[->, dashed, black, thick] (13.0,-4.8) -- (0.0,-4.8);
\coordinate[label=left:$t$] (A1) at (0.0,-4.8);
\coordinate[label=above:$B$] (A1) at (-0.1,-6.1);
\coordinate[label=above:$A$] (A1) at (2.6,-6.1);
\coordinate[label=above:$B$] (A1) at (3.7,-6.1);
\coordinate[label=above:$A$] (A1) at (8.1,-6.1);
\coordinate[label=above:$B$] (A1) at (12.3,-6.1);
\coordinate[label=above:$A$] (A1) at (13.1,-6.1);
\draw[->, black, thick] (2.3,-5.7) -- (0.3,-5.7);
\draw[->, black, thick] (3.4,-5.7) -- (2.9,-5.7);
\draw[->, black, thick] (7.8,-5.7) -- (4.0,-5.7);
\draw[->, black, thick] (12.0,-5.7) -- (8.4,-5.7);
\draw[->, black, thick] (12.9,-5.7) -- (12.6,-5.7);
\node (A) at (1.4,-6.4) {$I_+$};
\node (A) at (3.2,-6.4) {$I_-$};
\node (A) at (10.4,-6.4) {$I_-$};
\node (A) at (12.8,-6.4) {$I_+$};
\end{tikzpicture}
\end{center}
\caption{Illustration of trajectories and time-reversal. a) Coarse-grained two-state trajectory with sequence of transitions. b) Corresponding schematic Langevin trajectory divided into snippets between transitions. Their duration $\tau_i$ is indicated by curly brackets. The bijective mapping between $\mathcal{R}(\zeta_{I_+ \rightarrow I_+}^{\tau})$ and $\zeta_{I_- \rightarrow I_-}^{\tau}$ is possible by translating the dashed part of the trajectory in time. The starting point of the snippet of length $\tau_{2}$ is arbitrarily set to $t=0$. c) Time-reversed coarse-grained trajectory and sequence of transitions. The time-reversal changes the times at which the transitions are recorded as well as the duration of the snippets.}
\label{fig:Pfadgewichtbeweis}
\end{figure}
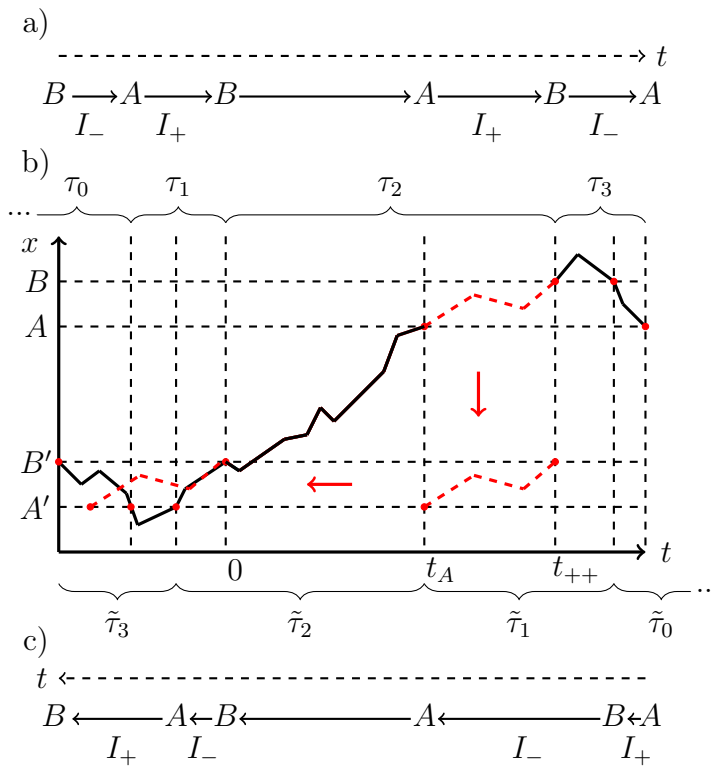

We assume that we cannot observe the full trajectory $\gamma = x(t)$ but only transitions $I_{+}$ and $I_-$. The effective trajectory $\Gamma$ equals a sequence of such transitions after certain waiting-times $\tau_{i}$. Therefore, the information gained through the measurement can be represented by tuples $(I_{\pm}, \tau_{i})$. For the section of trajectory shown in figure \ref{fig:Pfadgewichtbeweis}, we get
\begin{eqnarray}
\Gamma \; = \; ... \rightarrow (I_-,\tau_{0}) \rightarrow (I_+, \tau_{1}) \rightarrow (I_+, \tau_{2}) \rightarrow (I_-,\tau_{3}) \rightarrow ... \, .
\label{eq:sequnece}
\end{eqnarray}
%where the past and the future evolution of the system is irrelevant because of the Markov-property of transitions.
Such data retains all available information and thus suffice to evaluate the waiting-time distributions (\ref{eq:WT_CP}). 
The coarse-grained sequence $\Gamma$ can be realized on the microscopic level by many different trajectories $\gamma$, e.g., by
\begin{eqnarray}
\gamma \; = \; ... \rightarrow \zeta_{I_{-} \rightarrow I_{+}}^{\tau_{1}} \rightarrow \zeta_{I_{+} \rightarrow I_{+}}^{\tau_{2}} \rightarrow \zeta_{I_{+} \rightarrow I_{-}}^{\tau_{3}} \rightarrow ... \, ,
\end{eqnarray}
that consist of snippets $\zeta_{I \rightarrow J}^{\tau}$ of length $\tau$ connecting two transitions $I$ and $J$.
%Thus, a coarse-grained trajectory $\Gamma$ occurs with probability
%\begin{eqnarray}
%\mathcal{P}[\Gamma] = \sum_{\gamma \in \Gamma} \mathcal{P}[\gamma|I] \, ,
%\label{coarse_pathweight}
%\end{eqnarray}
%where we have summed up all microscopic trajectories $\gamma$ that are coarse-grained into $\Gamma$ starting at the endpoint $X_I$ of the first transition $I$. 
All contributing $\gamma$ share the same sequence of transitions and associated duration, i.e., the tuples in (\ref{eq:sequnece}).
We can express $\mathcal{P}[\Gamma]$ using the Markov property of transitions and sum over all snippets connecting the respective transitions to arrive at
\begin{eqnarray}
\mathcal{P}[\Gamma] = ... \sum_{\zeta \in \set{\zeta_{I_{-} \rightarrow I_{+}}^{\tau_{1}}}} \mathcal{P}[\zeta|A']  \sum_{\zeta \in \set{\zeta_{I_{+} \rightarrow I_{+}}^{\tau_{2}}}} \mathcal{P}[\zeta|B'] \sum_{\zeta \in \set{\zeta_{I_{+} \rightarrow I_{-}}^{\tau_{3}}}} \mathcal{P}[\zeta|B] \; ... 
\label{eq:Zerlegung1}
\end{eqnarray}
for the example shown in figure \ref{fig:Pfadgewichtbeweis}\,b).
Since the sums equal the waiting-times in (\ref{eq:WTDausPW}), the path weight (\ref{eq:Zerlegung1}) becomes
\begin{eqnarray}
\mathcal{P}[\Gamma] =  ... \; \psi_{I_{-} \rightarrow I_{+}}(\tau_{1}) \psi_{I_{+} \rightarrow I_{+}}(\tau_{2}) \, \psi_{I_{+} \rightarrow I_{-}}(\tau_{3}) \; ... \, .
\label{eq:zerlegung}
\end{eqnarray}
From the trajectory shown in figure \ref{fig:Pfadgewichtbeweis} it should be clear that, while all paths contributing to $\psi_{I_{\pm} \rightarrow I_{\pm}}(\tau)$ form closed loops, the ones contributing to $\psi_{I_{\pm} \rightarrow I_{\mp}}(\tau)$ do not because the latter begin and end at different milestones.

To compute the entropy produced by $\Gamma$, its time-reversed version 
\begin{eqnarray}
\tilde{\Gamma} \; = \; ... \rightarrow (I_+,\tilde{\tau}_{0}) \rightarrow (I_-, \tilde{\tau}_{1}) \rightarrow (I_-, \tilde{\tau}_{2}) \rightarrow (I_+, \tilde{\tau}_{3}) \rightarrow ... \, ,
\end{eqnarray}
is needed, for which we replace every $I_+$ transition with an $I_-$ transition and vice versa in the sequence (\ref{eq:sequnece}). In the time-reversed trajectory as shown in figure \ref{fig:Pfadgewichtbeweis}\,c), these transitions occur at different times, since they possess a finite duration. Hence, also the duration $\tau_{i}$ of a snippet changes to $\tilde{\tau}_i$ under time-reversal. Coarse-graining the time-reversed trajectory at the state-level thus leads to a phenomenon called kinetic hysteresis described in \cite{hartich_emergent_2021}, i.e., in this setup, time reversal and coarse-graining do not commute. Additionally, at the transition level, the duration of the snippets change and the transitions are measured at different absolute times.
The time-reversed version of a trajectory that undergoes two subsequent $I_+$ transitions will undergo two subsequent $I_-$ ones. However, the waiting-time gets altered, i.e., in the example $\tau_2 \neq \tilde{\tau}_2$.
In the forward trajectory the $I_+$ transition is finished at $t_{++} = \tau_2 $ after a preceding $I_+$ one, see the time axis. If we reverse the trajectory in time, this part of the trajectory does not correspond to a $I_- \rightarrow I_-$ sequence.
%, which demonstrates that time-reversal and coarse-graining in this setup do not commute. 

Thus, in order to compare the effect of time-reversal on the microscopic and on the coarse-grained level, we have to compare how the trajectories and waiting-time distributions are modified. Since it is known how the time-reversal operation $\mathcal{R}$ acts on the microscopic level, i.e. on trajectories $\gamma$ and snippets $\zeta_{I \rightarrow J}^{\tau}$, we can investigate whether every time-reversed snippet connecting two $I_+$ transitions can be mapped to a snippet with equal path weight connecting two $I_-$ transitions in the original dynamics, i.e,
\begin{eqnarray}
\mathcal{P}[\mathcal{R}(\zeta_{I_+ \rightarrow I_+}^{\tau})| \tilde{I}_+] = \mathcal{P}[\zeta_{I_ - \rightarrow I_-}^{\tau}| I_-]
\label{eq:toprove}
\end{eqnarray}
holds true.
In fact, this is possible and can be shown by rearranging every time-reversed $\zeta_{I_+ \rightarrow I_+}^{\tau}$ trajectory using the time-translational invariance of the path weight as illustrated in figure \ref{fig:Pfadgewichtbeweis}\,b). Consider the trajectory starting at $t=0$ and ending at $t_{++}$. The time-reversed version of it starts at $B$ and ends in $B'$. If we cut the dashed part, it does not matter whether we place it between $B'$ and $A'$ or between $B$ and $A$ because the states are physically equivalent and merely differ by $L$. In addition, the dashed path can be translated in time and thus added at the end of the time-reversed trajectory at state $B'$. This leads to a modified trajectory that equals a snippet connecting two $I_-$ transitions thus confirming (\ref{eq:toprove}). The length of the original snippet and the modified one coincide and they both complete the cycle once. As a consequence,
\begin{eqnarray}
\psi_{\tilde{I}_{+} \rightarrow \tilde{I}_{+}}(\tau) = \sum_{\zeta \in \set{\zeta_{I_+ \rightarrow I_+}^{\tau}}} \mathcal{P}[\mathcal{R}(\zeta)| \tilde{I}_+] = \psi_{I_{-} \rightarrow I_{-}}(\tau) 
\end{eqnarray}
holds. The ratio of path weights between the two snippets that complete the cycle once in opposite direction obeys  
\begin{eqnarray}
\frac{\mathcal{P}[\zeta_{I_+ \rightarrow I_+}^{\tau}| I_+]}{\mathcal{P}[\zeta_{I_- \rightarrow I_-}^{\tau}| I_-]} = \exp \left[\int_0^{\tau} F(x(t)) \dot{x}(t) \rmd t \right] = e^{f L} = e^{\mathcal{A}}
\label{eq:aff_mic_Path}
\end{eqnarray}
similar to \cite{berezhkovskii_identity_2006}. Multiplying by the denominator of the left-hand side and summing over all snippets leads to
\begin{eqnarray}
\psi_{I_{+} \rightarrow I_{+}}(\tau) = e^{\mathcal{A}} \psi_{I_{-} \rightarrow I_{-}}(\tau) \, ,
\end{eqnarray}
which implies that the estimator (\ref{eq:estimator_1}) is time-independent and that it yields the full affinity, just as the estimator for discrete dynamics does \cite{van_der_meer_thermodynamic_2022}. %Strikingly, our reasoning merely relied on the Markovianity of transitions and time-independence of $F(x)$. 

In addition, we are now able to state a relation that only uses waiting-time statistics to infer the total entropy $\Delta S[\gamma]$ along a microscopic trajectory $\gamma$ and its mean rate $\sigma$. In terms of the path weight $\mathcal{P}[\gamma|x_{\text{i}}]$, the total entropy production along $\gamma$ is given by 
\begin{eqnarray}
\Delta S[\gamma] = \ln \left(\frac{\mathcal{P}[\gamma|x_{\text{i}}] \, p^{\text{s}}(x_{\text{i}})}{\mathcal{P}[\tilde{\gamma}|x_{\text{f}}] \, p^{\text{s}}(x_{\text{f}})} \right) 
\label{eq:EP}
\end{eqnarray}
in the stationary state. Here the distribution of the initial points $x_{\text{i}}$ and $x_{\text{f}}$ of $\gamma$ and $\tilde{\gamma}$, respectively, are taken from the steady state distribution $p^{\text{s}}(x)$. In the long-time limit, the contribution from the initial conditions becomes neglectable.
The entropy produced by one trajectory $\gamma$ can be calculated as the sum of log-ratios between probabilities of snippets. With (\ref{eq:aff_mic_Path}) we can replace the entropy produced by closed loops by (\ref{eq:estimator_1}). Hence, it suffices to count the times $N_{I_{\pm}}(t)$ two successive $I_+$ and $I_-$ transitions are observed on the coarse-grained level up to time $t$ to arrive at
\begin{eqnarray}
\Delta S[\gamma] = (N_{I_+ \rightarrow I_+}(t) -  N_{I_- \rightarrow I_-}(t)) \ln\left(\frac{\psi_{I_{+} \rightarrow I_{+}}(t)}{\psi_{I_{-} \rightarrow I_{-}}(t)} \right) \, .
\end{eqnarray}
%Note, that $\Delta S[\gamma]$ fluctuates because the $N_{I_{\pm} \rightarrow I_{\pm}}(t)$ depend on the individual realization of $\Gamma$. 
Thus, waiting-time statistics can be used to infer the whole distribution of entropy production.
In the long-time-limit $I_{\pm} \rightarrow I_{\mp}$ do not contribute, since $N_{I_+ \rightarrow I_-}(t) - N_{I_- \rightarrow I_+}(t)$ is equal to $0, \pm 1$. The mean entropy production rate $\sigma$ then follows from (\ref{eq:EPR}) with
\begin{eqnarray}
j^{\text{s}} = \lim_{t \rightarrow \infty} \frac{N_{I_+ \rightarrow I_+}(t) -  N_{I_- \rightarrow I_-}(t)}{t} \, .
\end{eqnarray}

%Taking the time-derivative and using that the estimator (\ref{eq:estimator_1}) is constant, it suffices to argue that the stationary current can be represented by
%\begin{eqnarray}
%j^{\text{s}} = \nu_{I_+ \rightarrow I_+} - \nu_{I_- \rightarrow I_-}
%\end{eqnarray}
%with the rate of occurrence 
%\begin{eqnarray}
%\nu_{I_{\pm} \rightarrow I_{\pm}} = \frac{\partial_t \langle N_{I_{\pm} \rightarrow I_{\pm}}(t) \rangle}{\partial_t \langle N_{I_{\pm}}(t) \rangle}
%\end{eqnarray}
%of two subsequent $I_{\pm}$ transitions that replaces the stationary current in (\ref{eq:EPR}). In addition to the number of subsequent transitions, the total numbers $N_{I_{\pm}}(t)$ of $I_{\pm}$ transitions is needed for normalization and the brackets $\langle ... \rangle$ stand for an average over many realizations or one long trajectory in the steady state.

%\begin{table}
%\begin{center}
%\caption{Properties of transitions in discrete and continuous Markovian systems. (WT = waiting-times)}
%\label{tab:transitions}
%\begin{tabular}{ |c||c|c|c| } 
% \hline
% \hline
%  & duration & Markovian & time-reversal  \\ 
% \hline
% discrete & zero & yes & $\pm \rightarrow \mp$, same WT\\ 
% continuous & finite & yes & $\pm \rightarrow \mp$, other WT\\ 
% \hline
%\end{tabular}
%\end{center}
%\end{table} 

\textit{Summarizing perspective.--}
We have identified a Markovian event for one-dimensional, overdamped Langevin dynamics that can be used to infer the affinity in a uni-cyclic system by analyzing waiting-times between successive events. Transitions correspond to passages between two milestones with a certain orientation that is detected using an auxiliary state in between. Just like for discrete dynamics, the key property of events that here yield an exact estimator is Markovianity \cite{van_der_meer_time-resolved_2023}. Furthermore, we obtain the distribution of entropy production from waiting-time distributions.

The presented formalism can easily be applied to arbitrary graphs consisting of connected, one-dimensional continuous paths like the ones described in \cite{hartich_emergent_2021}. If we define transitions as passages between nodes of the network in the same manner as for the uni-cyclic system, the results obtained in \cite{van_der_meer_thermodynamic_2022} for multi-cyclic discrete systems hold true. To be specific, the estimator (\ref{eq:EPR}) then becomes time-dependent and is bounded by the smallest affinity and the largest one of all cycles that contain the observed transition. Moreover, our approach simplifies the theoretic framework used in \cite{hartich_emergent_2021} because only the Markovian property of the milestones is required to adapt the proofs based on snippets. It further underlines that milestoning always bounds the entropy production if reaching the milestone is a Markovian event \cite{van_der_meer_time-resolved_2023}. 

In principle, this approach is not limited to one-dimensional systems. For a two-dimensional one like a particle driven along a torus it would be interesting to exploit the definition of transitions between milestones that in this setup could correspond to closed curves that cannot be contracted to a point. However, since the Markov property only applies to points and not to curves, hitting a curve is a non-Markovian event. Thus, the discussion from \cite{van_der_meer_time-resolved_2023} applies and we expect that the estimator (\ref{eq:estimator_1}) does not yield a bound on the full entropy production. 

Another open question is the application to underdamped Langevin dynamics for which the distinction between forward and backward transitions does not require the auxiliary point $C$ if positive and negative velocities at one single milestone could be discerned. However, without access to the precise value of the velocity, registering such an event is still non-Markovian \cite{noauthor_fluctuating_nodate}, which means that a bound on entropy production cannot be expected, in general.

\ack
We thank Jann van der Meer for stimulating discussions.

\section*{References}
\bibliographystyle{unsrt}
\bibliography{Paper}

\end{document}